\documentclass{sig-alternate-10pt}
\usepackage{amssymb}
\usepackage{algpseudocode}
\usepackage{algorithm}
\usepackage{multirow}
\usepackage{amssymb}
\usepackage{array}
\usepackage{graphicx}
\usepackage{cite}
\usepackage{textcomp}
\usepackage{color}
\usepackage{booktabs}
\usepackage{dcolumn}
\usepackage{fixltx2e}

\begin{document}

\title{MSPlayer: Multi-Source and multi-Path LeverAged YoutubER}
\numberofauthors{4}
\author{
\alignauthor
Yung-Chih Chen\\
      \affaddr{UMass Amherst}\\
      \affaddr{USA}\\
       \email{yungchih@cs.umass.edu}\\
\alignauthor
Don Towsley\\
        \affaddr{UMass Amherst}\\
      \affaddr{USA}\\
       \email{towsley@cs.umass.edu}\\
\alignauthor
Ramin Khalili\\
      \affaddr{T-labs/TU-Berlin}\\
      \affaddr{Germany}\\
       \email{ramin@inet.tu-berlin.de}
}

\maketitle

\begin{abstract}
Online video streaming through mobile devices has become extremely popular
nowadays. YouTube, for example, reported that the percentage of its traffic streaming to mobile devices has soared from 6\% to more than 40\% over the past two
years. Moreover, people are constantly seeking to stream high quality video for better experience while often suffering from limited bandwidth. 
Thanks to the rapid deployment of content
delivery networks (CDNs), popular videos are now replicated at
different sites, and users can stream videos from close-by locations
with low latencies. As mobile devices nowadays are equipped with multiple wireless interfaces (e.g., WiFi and 3G/4G), aggregating bandwidth for high definition video streaming has become possible. 

We propose a client-based video streaming solution, MSPlayer, that takes advantage of multiple video sources as well as multiple network paths through different interfaces. MSPlayer reduces start-up latency and provides high quality video streaming and robust data transport in mobile scenarios. We experimentally demonstrate our solution on a testbed and through the YouTube video service.
\end{abstract}

\section{Introduction}
\label{sec:intro}

With the high demand for online video streaming, video content providers are offering better technology to satisfy customers' desires for streaming high quality videos. However, streaming video to a user nowadays still encounters the following issues. First, people from time to time experience insufficient bandwidth. Research has shown that viewers are not patient enough to wait for a start-up delay longer than a few seconds \cite{krishnan2012video} and video quality has a huge impact on user engagement \cite{dobrian2011understanding}. Also, connections to a particular network can break down temporarily due to mobility and re-establishing a connection introduces additional delays. Last, as network bandwidth is highly variable, the commercial video players have experimented with video rate adaption, which in turn results in performance issues such as variable video quality, unfairness to other players, and low bandwidth utilization\cite{jiang2012improving,akhshabi2012happens,huang2012confused,akhshabi2011experimental, mok2012qdash}.    
 
As mobile devices are now equipped with multiple wireless interfaces connected to different networks (WiFi or cellular 3G/4G), one possible solution is to use multi-path TCP (MPTCP) \cite{RFC6824MPTCP}. However, since MPTCP requires kernel modifications at both the client and server sides\cite{raiciu2012hard}, and many network operators do not allow MPTCP traffic to pass their middleboxes \cite{Honda2011, hesmans2013tcp}, it has been slow to deploy MPTCP globally. Furthermore, although MPTCP provides a means for balancing loads over different paths to a single server, it does not utilize source diversity in the networks in order to facilitate content delivery. Therefore, it is urgent and necessary to develop a solution to stream high quality videos to end users without overloading the video servers. Thus, we ask the following question: 
\emph{Is it possible to leverage both path diversity and source diversity to provide robust video delivery and to reduce video start-up latency?}

In this paper, we take a first step to answering this question. We show that one can utilize both of the available WiFi and 4G interfaces simultaneously to aggregate bandwidth for higher quality video streaming. The video streaming solution does not require modifications in the kernel stacks and is not hindered by network middleboxes. We then instantiate these designs in our YouTube player, MSPlayer. By investigating the YouTube service architecture and its streaming mechanisms, we further demonstrate how to leverage the existence of multiple video sources in different networks at the same time. We then experimentally evaluate the performance of different MSPlayer schedulers as well as the performance of MSPlayer through the YouTube service.  

The remainder of this paper is organized as follows: Sec.~\ref{sec:design} introduces the design principles of MSPlayer. We overview the architecture of MSPlayer in Sec.~\ref{sec:overview}. Implementation details of MSPlayer are described in Sec.~\ref{sec:implementation} and we evaluate MSPlayer's performance in our testbed and over YouTube infrastructure in Sec.~\ref{sec:testbed_evaluation} and \ref{sec:evaluation}. We discuss future work and conclude this paper in Sec.~\ref{sec:future}.

\newpage
\section{Design Principles} 
\label{sec:design}

In this paper, we present MSPlayer, a client-based approach for video streaming that requires no changes in either the server or the client's kernel stacks. It leverages diversity in the network and performs load balancing at the client side. MSPlayer also supports user mobility and provides robust data transport. In order to be fair to other TCP users, MSPlayer limits the number of paths to two (one over WiFi and one over 3G/4G) and  leverages HTTP range requests to stream videos. It has the following design features. 
\newline
\newline {\bf Just-in-time with High Quality:\ }
Since viewers often prematurely stop watching videos \cite{finamore2011youtube, krishnan2012video}, streaming the entire video to a viewer at once can result in waste of bandwidth and network resources. This, along with the rise of adaptive streaming over HTTP \cite{stockhammer2011dynamic}, has drawn attention to \emph{just-in-time} video delivery, which has been exploited by most large scale video streaming services such as YouTube, Netflix, and Hulu. A video is partitioned into many small file segments called \emph{video chunks}. The video server maintains multiple profiles of the same video for different bitrates and video quality levels. Clients then \emph {periodically} request video chunks and adapt video bitrates. 

Just-in-time video delivery avoids a waste of resources if a user drops the video during its playback. Dynamically adapting video bitrates, however, results in performance problems such as low link utilization \cite{akhshabi2012happens}, unfairness to other TCP users \cite{huang2012confused, akhshabi2012happens}, and unstable video quality \cite{akhshabi2011experimental, mok2012qdash}. In our design, we share the same just-in-time concept for video delivery. However, we do not investigate rate adaption and focus on how to stream videos to users with a \emph{fixed bitrate} by exploiting \emph{network diversity}.
\newline
\newline{\bf Robust Data Transport:\ }
When a mobile user streams a video, his connection (mostly WiFi) can break and the downloaded video will thus be abandoned. In order to resume the video, the user then needs to switch to another available interface with connectivity to another network, establish a new TCP connection, and move/skip to the break point. In the worst case, one will have to wait until reaching the next WiFi hotspot and repeat the entire process mentioned above.  

One possible solution is to use Multi-Path TCP (MPTCP) \cite{RFC6824MPTCP}, which has been standardized by the IETF, aimed at providing robust data transport. However, MPTCP still faces several deployment challenges: First, MPTCP requires kernel modifications at \emph{both} the server and client sides \cite{raiciu2012hard}. Second, it relies on the TCP option field to exchange path and interface information. In the later case, research has shown that MPTCP suffers significantly from network middleboxes as they very often strip away unknown options \cite{Honda2011, hesmans2013tcp}, forcing MPTCP connections to fall back to legacy single-path TCP. In our measurements, two out of three major US cellular carriers do not allow MPTCP traffic to pass through the default HTTP 80 port, which is a potential problem for video streaming to popular sites such as YouTube or Netflix. 

We design a client-based multi-path solution to provide robust data transport for high quality video streaming. Furthermore, each path runs legacy TCP and is therefore guaranteed to successfully pass network middleboxes.  
\newline\newline
\newline{\bf Content Source Diversity:\ }
Current MPTCP design \cite{RFC6824MPTCP} and other similar approaches such as \cite{bui2013greenbag}, only allow flows or paths to be established between a client and \emph{a single server}. If the current YouTube infrastructure were to support MPTCP, users streaming videos from one server with high aggregate bandwidth through multiple paths could quickly incur server demand surges. This high demand, particularly for high quality videos, can overload the server and congest shared bottleneck links. The outcome of this can directly or implicitly affect other viewers' experience.

As popular content is now replicated at multiple locations or data centers, content delivery networks (CDNs) are responsible for handling video replicas and delivering videos across different data centers for large scale video streaming services such as YouTube, Netflix, and Hulu \cite{adhikari2012vivisecting, adhikari2012unreeling}. As part of our design is to provide robustness, MSPlayer, at the initial phase, collects a list of YouTube servers' addresses in each network exploited. If a server in a network fails or is overloaded, MSPlayer switches to another server in that network and resumes video streaming. Other proposals, such as \cite{gouache2011distributed}, aim to emulate the use of multiple paths in a controlled environment by setting up multiple connections to the servers connected by a switch with only one \emph{single interface}. Although this approach can potentially distribute the load among the connected servers, having multiple connections over one interface could quickly saturate the bottleneck link.  

As wireless interfaces are associated with different networks, MSPlayer requests partial content from \emph{video servers in all networks} simultaneously to avoid overwhelming particular video servers and to balance the load across the servers. In this work, we use Google's public DNS service to resolve the IP addresses of YouTube servers. 
\newline\newline {\bf Chunk Scheduler:\ }
MSPlayer relies on range requests to retrieve video chunks over different paths. As making a range request incurs additional overhead (packets start to arrive one RTT after the request is sent) and different paths usually exhibit diverse latencies \cite{chen2013measurement}, how to schedule chunks over different paths efficiently is challenging. Therefore, it is desirable to have a good scheduler that estimates path quality over time and efficiently assigns chunks to each path. 

To satisfy just-in-time video delivery, the scheduler pauses chunk retrieval when the playout buffer is full and resumes chunk requesting when the amount of buffered video falls below a certain level (that is referred to as periodic downloading or ON/OFF cycles \cite{rao2011network}). To reduce memory usage of out-of-order chunks from different paths, the MSPlayer scheduler attempts to complete the transfer of a chunk over each path at the same time, and allows \emph{at most one} out-of-order chunk to be stored.


\section{MSPlayer Overview}
\label{sec:overview}
We now overview the MSPlayer architecture. We first describe how YouTube video streaming works and the just-in-time video delivery, followed by descriptions of the MSPlayer's design components: multi-source, multi-path, and chuck scheduler. 

\subsection{YouTube Video Streaming}
\label{sec:background}
\begin{figure*}
\begin{center}
\includegraphics[width=0.7\textwidth]{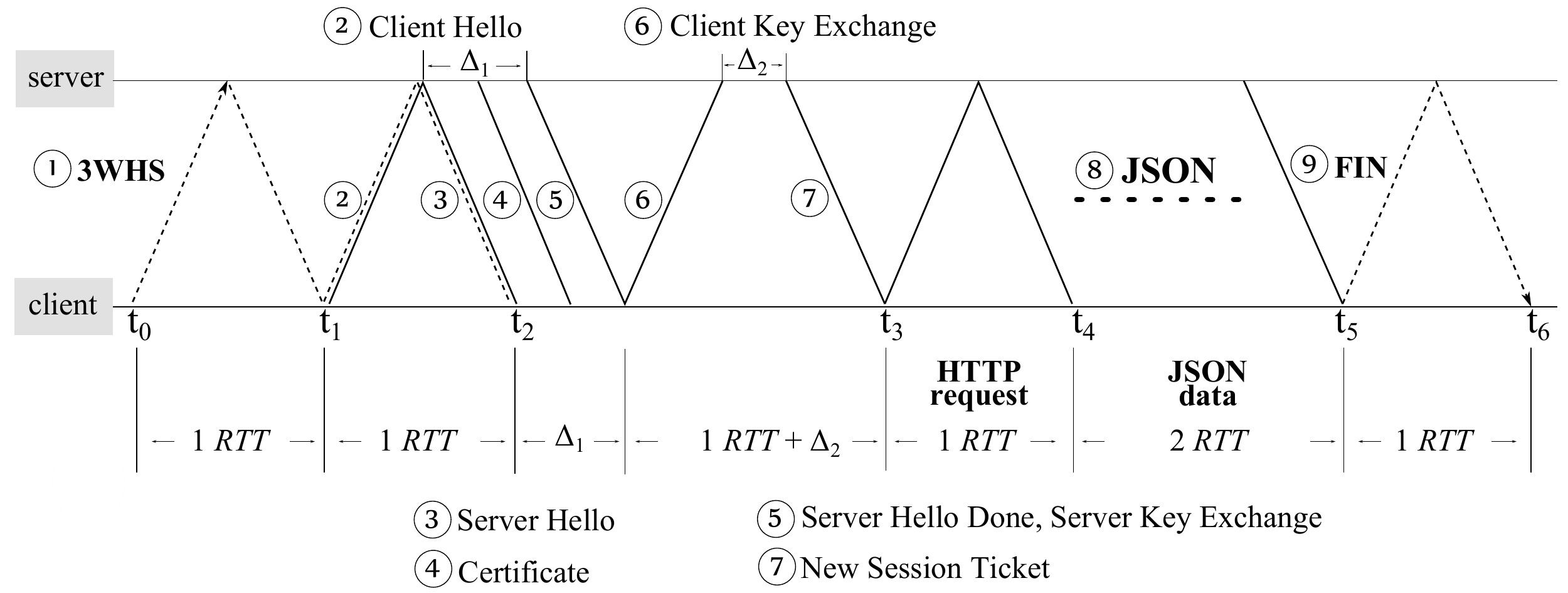} %
\end{center}
\vspace{-.4cm}
\caption{HTTPS connection to YouTube web server: retrieving JSON objects of video information.}
\label{fig:HTTPS}
\end{figure*}


People usually go to the YouTube website and choose a video to watch, or just click on an URL of the following form \texttt{http://www.youtube.com/watch?v=qjT4T2gU9sM} on a web page. Users can then watch the video through their browsers using a built-in Adobe Flash player \cite{AdobeHTTPdynamicstreaming}. Each YouTube video is identified by an 11-literal video ID after \texttt{watch?v=} in the URL \cite{adhikari2012vivisecting}. 

With this URL, the video player (e.g., Adobe Flash) first performs a DNS lookup to resolve the IP address of the domain name \texttt{www.youtube.com} and the user's video request is then directed to one of YouTube's web proxy servers. The YouTube web proxy server processes the request and returns the related video information and a new URL to the user in JavaScript Object Notation (JSON) format, indicating where the associated and available YouTube video servers are. The player then establishes another connection to one of the dedicated video servers and starts to stream the YouTube video using HTTP range requests. 

The streaming process starts with a \emph{pre-buffering phase} followed by a periodic \emph{re-buffering phase} \cite{rao2011network}. The pre-buffering phase takes place at the beginning of the streaming, aimed at retrieving enough video data in the playout buffer. When the amount of video content in the playout buffer falls below a certain level, the player enters the re-buffering phase, making new requests to refill the playout buffer. This periodic re-buffering repeats until the video is completely watched or dropped.

\subsection{Multi-Source and Multi-Path}
\label{sec2:msmp}

%

Before describing our scheme with multiple sources and multiple paths, we first describe how each path establishes a connection to the YouTube web proxy server and the associated video server. Fig.~\ref{fig:HTTPS} illustrates a flow diagram when a user contacts YouTube's web proxy server to retrieve video information. The connection starts with a TCP 3-way handshake (3WHS). Afterwards, the client initiates a secure connection handshake message at time $t_1$. It takes the server times $\Delta_1$ and $\Delta_2$ to verify the key and complete the key exchange process. The first HTTP request is made at time $t_3$, and the first JSON packet from the web proxy server arrives at $t_4$. Note that these JSON packets are delivered within two round trips (slightly less than 20 packets)\footnote{As of July 2014, YouTube has applied algorithms to encode copyrighted video signatures. Since these signatures are needed to contact the video servers, for copyrighted videos, an additional operation is required to fetch the video web page containing a decoder to decipher the video signature.}, and the secure connection ends at $t_5$ followed by a TCP FIN. 

If we denote by $R_1$ and $R_2$ the RTTs of the first and the second paths, and by $\theta=R_2/R_1$ the RTT ratio (assuming $R_1\leq R_2$, i.e., the first path is a fast path), it takes time $\eta_i=4R_i+\Delta_1+\Delta_2$ to establish a secure HTTP connection over path $i$, and time $\psi_i=6R_i+\Delta_1+\Delta_2$ to receive complete video information before contacting the video server. If the YouTube's web proxy server is close to the video server, and both servers have similar capability of key verification, it takes approximately time $\pi_i\approx\psi_i+\eta_i$ for path $i$ to receive the first video packet from the video server. As part of chunk scheduling in MSPlayer, the first path is designed to contact its video server as soon as its IP address is decoded, without waiting for the second path to finish its decoding process. Chunks are scheduled over the first path before the second path becomes available. Therefore, before the second path starts to retrieve video packets from its associated video server, the first path will have received video packets for a duration of $\pi_2-\pi_1 \approx 10\cdot (R_2-R_1)= 10\cdot(\theta-1)R_1$. 

In MSPlayer, the processes of fetching video chunks over each path are executed by independent threads, which are under the management of the chunk scheduler (described in the next section).


\subsection{Chunk Scheduler}
\label{sec2:scheduler}
In order to reduce out-of-order delay for video streaming, and to reduce memory usage to store out-of-order chunks, our design is to schedule chunks (of different sizes) over both paths so that chunk transfers \emph {complete} at roughly the same time. 

To optimize video streaming performance with MSPlayer, chunk size selection for each path is critical and should be adapted over time in response to network dynamics. A previous measurement study shows that YouTube players, such as Adobe Flash or HTML5, use 64 KB and 256 KB as their default chunk sizes, while Netflix player (silverlight) uses larger chunk sizes that range from 2 MB to 4 MB \cite{rao2011network}. Since different mobile devices have pre-buffering periods of different lengths (ranging from 20 seconds to 1 minute) \cite{raiciu2012hard}, we also investigate the performance of different schedulers when applying different chunk sizes and pre-buffering periods. 

We denote by $S_i(t)$ the chunk size of path $i$ at time $t$, by $B$ the base  chunk size, and by $T_i$ the time required to download chunk $S_i(t)$. The estimated throughput to download $S_i(t)$ is denoted by $w_i(t) = S_i(t)/T_i$. 

We first showcase a baseline scheduler called \emph{Ratio} and then propose two different chunk size schedulers that adjust chunk sizes according to network bandwidth changes, namely the exponential weighted moving average (\emph{EWMA}) and \emph{Harmonic}. 
MSPlayer's chunk size selection should adapt to path quality variations over time, and the bandwidth estimator of MSPlayer thus plays a critical role in the chunk size selection process. In this paper, we label the chunk scheduler according to the bandwidth estimator used. We compare and evaluate the performance of these three schedulers in our testbed. 
\newline\newline { \bf Baseline Scheduler:\ }
Suppose $w_i(t)\leq w_{1-i}(t)$, the baseline \emph{Ratio} scheduler assigns a fixed chunk size to the path with lower throughput such that $S_i(t+1)=B$ and adjusts the chunk size of the path with higher throughput based on throughput ratio (i.e., $S_{1-i}(t+1)= w_{1-i}(t)/w_i(t)\cdot B$ and $i=0,1$ labels the first and the second path).  
\newline\newline { \bf Dynamic Chunk Adjustment Scheduler:\ }
\begin{algorithm}[t]
\caption{Dynamic chunk size adjustment}
\label{alg:adjustment}
\begin{algorithmic}[1]
\small
\Procedure{DCSA}{$i, \hat{w_0},\hat{w_1}, w_i, \delta, B$}  \Comment{$i=$ 0,1}
  	  \If {$\hat{w}_i$ not available}			
  	  	\State $S_i\gets B$                \Comment{initial chunk size}
   	  \ElsIf {$\hat{w_i} < \hat{w}_{1-i}$}  \Comment{slow path}
   	  	\If {$w_i>(1+\delta)\hat{w}_i$}
   	  		\State $S_i\gets 2\cdot S_i$
   	  	\ElsIf{$w_i<(1-\delta)\hat{w}_i$}   
   	  		\State $S_i\gets \max\{\lceil S_i/2 \rceil$, \textit{16KB}$\}$
   	  	\Else
   	  		\State $S_i$ unchanged
   	  	\EndIf
   	  	
   	  \Else \Comment{fast path}
   	  	\State $\gamma =\lceil \hat{w_i}/\hat{w}_{1-i} \rceil$
   	  	\State $S_i \gets \gamma \cdot S_{1-i}$
   	  \EndIf
   \State \textbf{return} $S_i$\Comment{final chunk size}
\EndProcedure
\end{algorithmic}
\end{algorithm}
When path bandwidth estimates are available, the chunk size of each path is adjusted according to Alg.~\ref{alg:adjustment}. We denote by $\delta$ the  throughput variation parameter in Alg.~\ref{alg:adjustment}. If the current bandwidth measurement of the slow path is ($1+\delta$) times larger than the estimated value, the chunk size is doubled. Similarly, if the current value is ($1-\delta$) times worse than the estimated value, the chunk size is halved. The chunk size of the fast path is adjusted based on the throughput ratio.

Here we focus on two bandwidth estimators: exponential weighted moving average (\emph{EWMA}) and harmonic mean (\emph{Harmonic}).
The weighted moving average is defined as: 
\begin{equation}
\hat{w_i}(t+1) = \alpha \cdot \hat{w_i}(t) + (1-\alpha)\cdot w_i(t).
\end{equation}
Due to the space constraint, in this paper we only report the results of $\alpha=0.9$ (details see next section).

As network bandwidth can vary quickly, extreme measurement values can bias our estimated results, following we introduce another bandwidth estimator called harmonic mean. The benefit to estimating path bandwidth by harmonic mean is that it tends to mitigate the impact of large outliers due to network variation
\cite{jiang2012improving}.

Given a series of bandwidth measurements, $w_i(t)$, where $t=${\small$0,1,2,\cdots,n-1$} and $w_i(t)>0$, the harmonic mean is $\hat{w}_i(n)=n/\sum_{t=0}^{n-1} \frac{1}{w_i(t)}$. A key factor of computing harmonic mean is to maintain a number of past measurements \cite{jiang2012improving}. However, to reduce memory usage and computational cost, one can compute the current harmonic mean without maintaining all previous states. Statistics from the past can be recovered simply by recording an additional parameter, $n$, the total number of past measurements. The harmonic mean can be updated with the most recent measurement of path $i$, $w_i(n)$, and the previous harmonic mean $\hat{w}_i(n)$. That is,
\begin{equation}
\hat{w}_i(n+1)=\frac{n+1}{\sum_{t=0}^{n} \frac{1}{w_i(t)}}= \frac{n+1}{\frac{n}{\hat{w}_i(n)} + \frac{1}{w_i(n+1)}}.
\end{equation}  

In the following sections, we describe MSPlayer implementation details and evaluate MSPlayer's performance.


\section{MSPlayer Implementation}
\label{sec:implementation}

In order to exploit both available wireless interfaces simultaneously, we pass additional interface information to the socket API to bind each interface to an IP address and packets can thus be scheduled to a desired interface. Moreover, we configure an independent routing table for each interface so that when a source IP address is specified, instead of using the default interface and gateway, the desired interface and gateway are used. 
Since video players can access YouTube videos through Google's Data APIs \cite{google}, MSPlayer is developed to leverage source and path diversity in the network for YouTube video streaming by interacting with Google APIs.\footnote{As YouTube's client libraries are mostly in web languages, MSPlayer, however is programmed in python rather than JavaScript or PHP.} 

First, when the desired video object is chosen, the player contacts the web proxy server with the URL containing the 11-literal video ID. The web proxy server then authenticates the user (player type and/or the user account) with \texttt{OAuth 2.0} and verifies the video operations requested by the user \cite{google}. When the requested operations are granted, the web proxy server resolves the user's public IP address and check to see which video server is suitable and available to this user based on YouTube's server selection mechanism \cite{adhikari2011you}. Afterwards, the web proxy server generates an access token (valid for an hour) that matches the video server's IP address as well as the operations requested. 

The web proxy server then encodes the token, together with the user's public IP address and the video's information (i.e., available video formats and quality, title, author, file size, video server domain names, $\ldots$ , etc) in JavaScript Object Notation (JSON) format and returns these objects to the user through the requested interface. MSPlayer then decodes the JSON objects received on each interface and synthesizes a new URL (with the required information, video server address, and a valid token) to contact the corresponding video server in the associated network. Video content is then retrieved by making HTTP range requests to different video servers with \emph{persistent connections} through both interfaces. Note that YouTube has been gradually replacing insecure HTTP connections with secure ones. To be compatible to the current and future YouTube's data service, we follow YouTube's latest HTTPS connection policy with both web proxy servers and video servers.

As part of the \emph{just-in-time} video delivery principle, MSPlayer uses the following streaming strategy similar to commercial YouTube players such as Adobe Flash player or HTML5: a pre-buffering phase followed by a steady-state re-buffering phase \cite{rao2011network}. MSPlayer leaves the pre-buffering phase when more than 40-second video data is received. It then consumes the video data until the playout buffer contains less than 10-second video. MSPlayer resumes requesting chunks from both YouTube servers and refills the playout buffer until 20 seconds of video data are retrieved. 


\section{Testbed Experimentation}
\label{sec:testbed_evaluation}
We first evaluate the performance of each design component of MSPlayer on a testbed in a controlled environment that emulates YouTube's video streaming mechanisms. The final performance evaluation is carried out on the YouTube infrastructure and service (see Sec. \ref{sec:evaluation}). Two types of servers are emulated in our testbed: web proxy servers (responsible for authentication and video object information delivery) and video servers. Both types of servers use the standard Linux 3.5 kernel with \texttt{CUBIC} congestion control \cite{ghobadi2012trickle} with Apache service.
Each type of server is hosted in two different UMass subnets for source diversity.   

The client running MSPlayer for video streaming is a Lenovo
X220 laptop equipped with a built-in 802.11 a/b/g/n WiFi interface connecting to a home WiFi network and an LTE dongle connecting to one of the major US cellular carriers. Video requests are sent over both interfaces simultaneously to two different YouTube video web proxy servers. Upon receiving packets from the web server, MSPlayer decodes the associated JSON objects (with a pre-loaded video server's IP address in our testbed) and fetches video chunks from the video servers. In our testbed, the videos are pre-downloaded in the servers from YouTube with MP4 format of HD (720p) video quality and 44,100 Hz audio quality.
\subsection{Multi-Source and Multi-Path}
\label{sec2:msmp_evalute}

\begin{figure}
\raggedright
\includegraphics[scale=0.5]{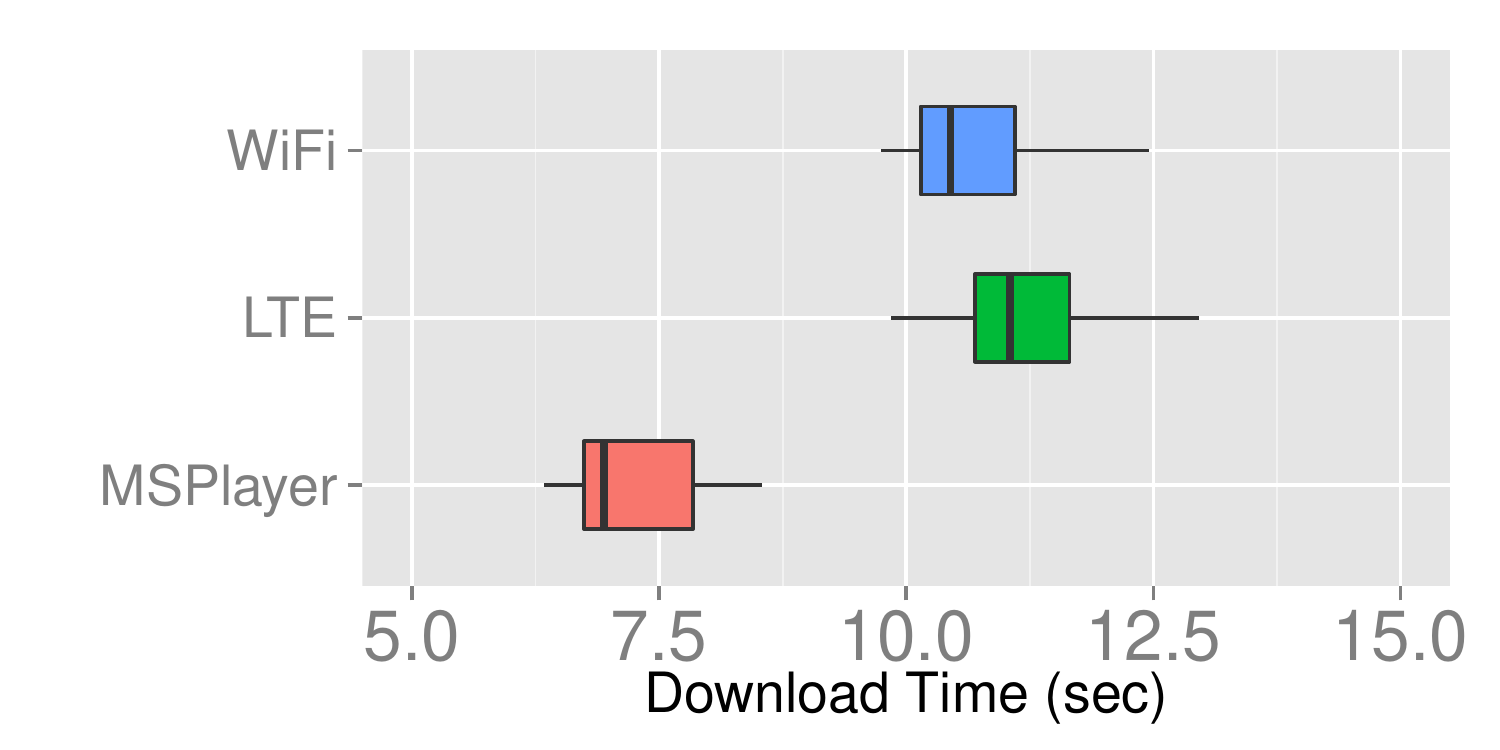}
\caption{Comparison of MSPlayer with WiFi and LTE for 40-sec pre-buffering (emulated).}
\label{fig:sp_msplayer}
\end{figure}

Fig.~\ref{fig:sp_msplayer} demonstrates the initial video pre-buffering download time using single-path WiFi, single-path LTE, and MSPlayer for HD videos in our emulated testbed. Note that a 40-sec pre-buffering period is presented here as this is YouTube servers' default pre-buffering size for Flash videos \cite{rao2011network}. The median download time of MSPlayer is 6.9 seconds while that of the best single-path over WiFi is 10.9 seconds, a 37\% delay time reduction in the pre-buffering phase. 

As MSPlayer leverages multiple video sources and interfaces/paths, how packets are scheduled through each path to each server can significantly affect the performance. The MSPlayer results in Fig. \ref{fig:sp_msplayer} are based on the Ratio scheduler with initial chunk size 1 MB. Next, we will investigate different MSPlayer schedulers and evaluate their performance. 


\subsection{Chunk Scheduler}
\label{sec2:secduler_evaluate}
We examine the performance of the following three schedulers: \emph{Harmonic}, \emph{EWMA}, and \emph{Ratio} (the baseline). We first examine the download time for different pre-buffering durations (for 20/40/60 seconds). For each pre-buffering duration, we further inspect each scheduler's performance with respect to different initial chunk sizes (from 16 KB to 1 MB). Throughout the experiments, we randomize the order in which the configurations are tested and repeat this 20 times over the course of 12 hours. Note that the throughput variation parameter $\delta=5\%$ and the \emph{EWMA} weight $\alpha=0.9$. 

\begin{figure}
\raggedright
\includegraphics[scale=0.5]{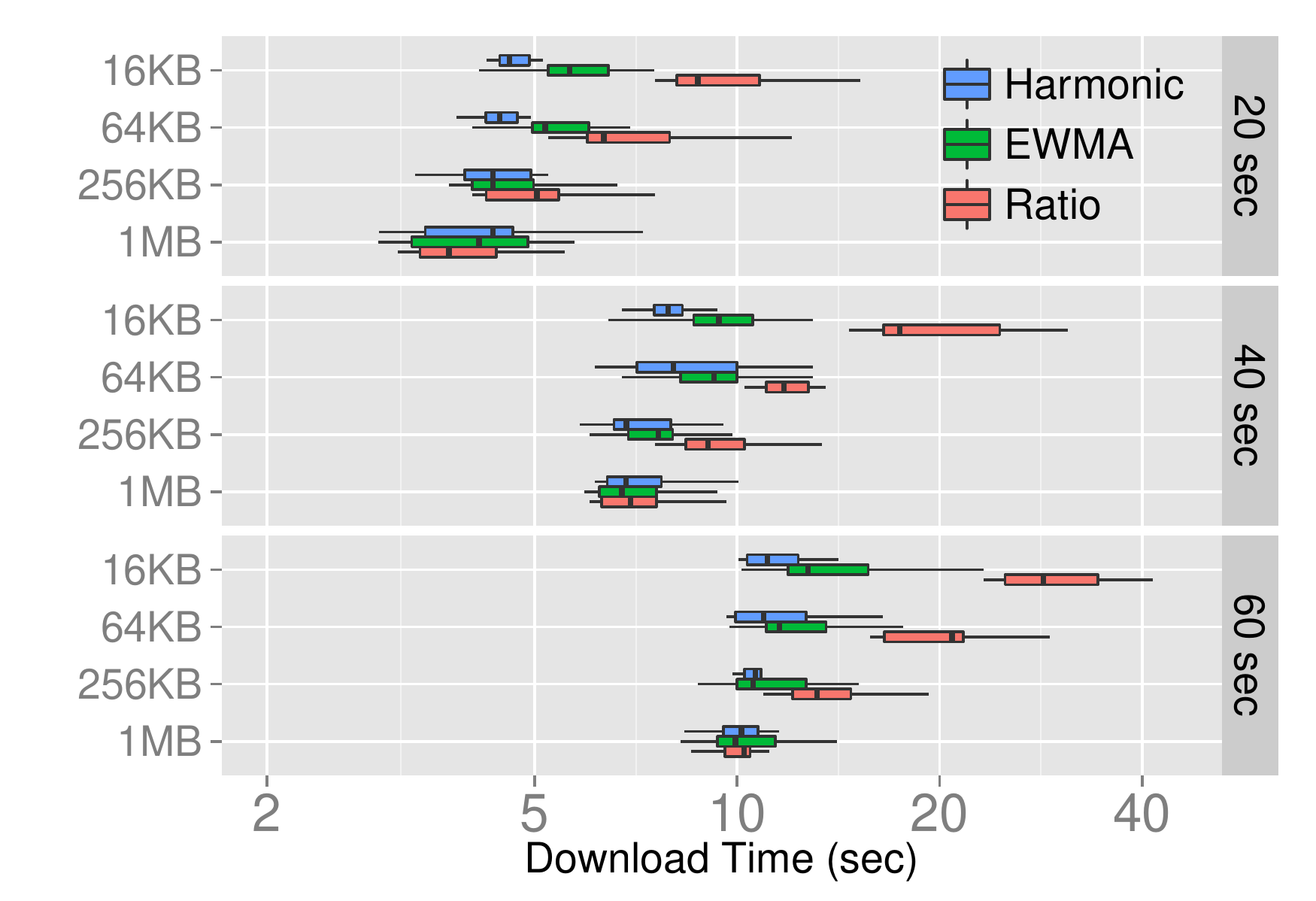}
\caption{Download times of three schedulers: Harmonic/EWMA/Ratio (top to down order) for different pre-buffering periods (right Y-axis) and initial unit chunk  sizes (left Y-axis).}
\label{fig:schedulers}
\end{figure}

As shown in Fig.~\ref{fig:schedulers}, for each pre-buffering duration, download time decreases as chunk size increases. For small chunk sizes, MSPlayer requires more range requests to accumulate the same amount of video in the pre-buffering phase. For larger chunk sizes, fewer requests are made and less overhead is included in the download time. 

The baseline scheduler does not perform well and exhibits higher variability as it fails to respond to bandwidth changes quickly. Dynamic chunk size adjustment schedulers (\emph{EWMA} and \emph{Harmonic}), on the other hand, vary path chunk sizes according to estimated bandwidth and exhibit better performance. More specifically, the scheduler using the harmonic mean estimator outperforms the others in most cases as this estimator is designed to mitigate large outliers such as large bursts. In our experiments, we use harmonic mean estimator as the default estimator in our scheduler. Since the performance of the harmonic mean scheduler is similar for chunk sizes 256 KB and 1 MB, we use a default initial chunk size of 256 KB as smaller chunk sizes are preferable to reduce network bursts \cite{ghobadi2012trickle}.

\section{Evaluation on YouTube Service}
\label{sec:evaluation}

\begin{figure}[t]
\raggedright
\includegraphics[scale=0.47]{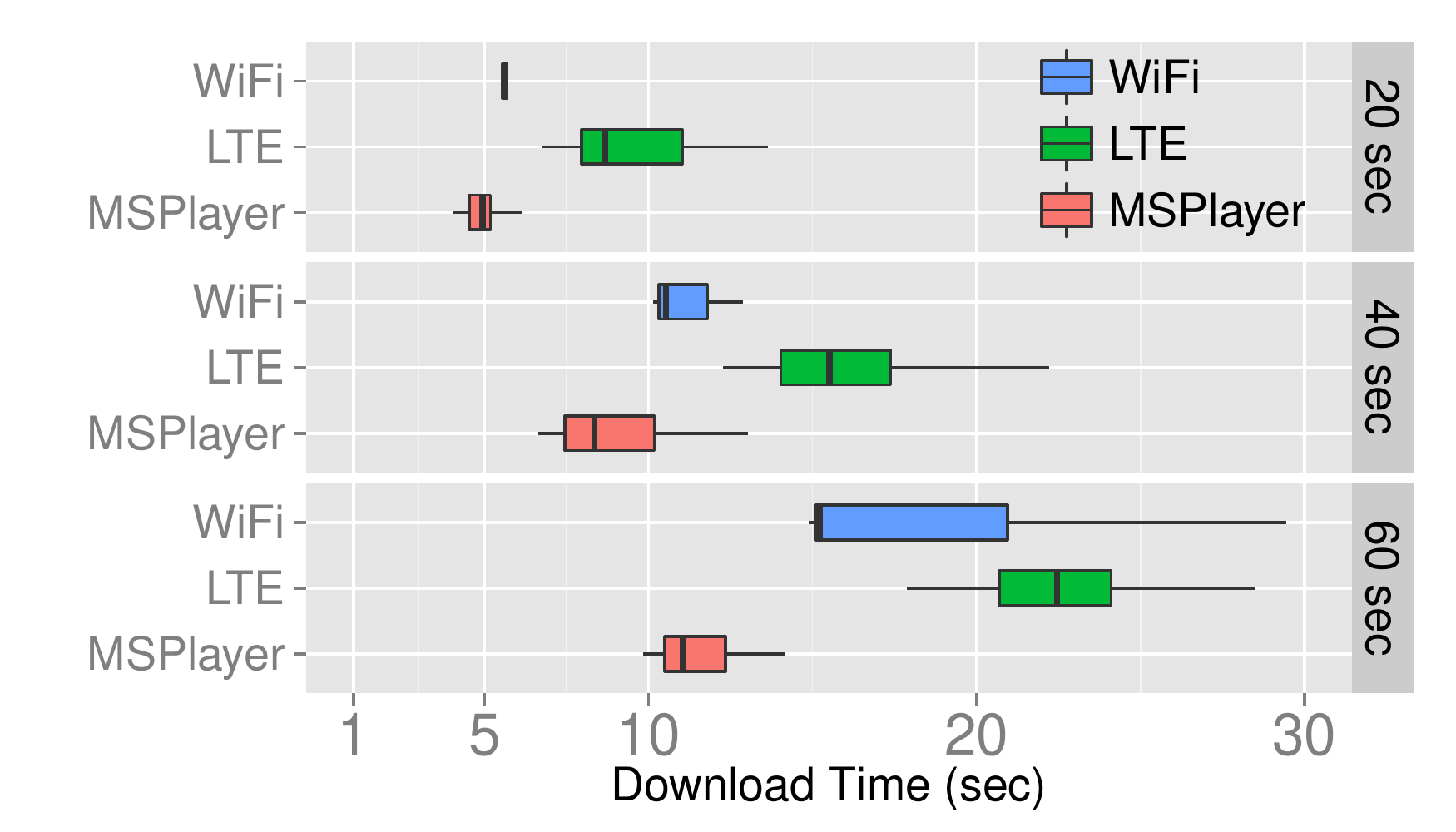}
\caption{Pre-buffering 20/40/60 second video for single-path WiFi, LTE, and MSPlayer on YouTube.}
\label{fig:pre_buf}
\end{figure}

We evaluate MSPlayer performance over the YouTube video infrastructure by comparing the download time of MSPlayer and the streaming schemes of the commercial YouTube players in both the \emph{pre-buffering phase} and the \emph{re-buffering phase}. We first look at the pre-buffering phase (where commercial players accumulate video data of a specified amount as one large chunk) and check to see how MSPlayer can reduce the start-up latency. Fig. \ref{fig:pre_buf} shows that MSPlayer outperforms both single-path TCP over WiFi and LTE for different specified amounts of pre-buffered video data. In comparison to the best single-path technology used, MSPlayer reduces start-up delay by 12\%, 21\%, 28\% for 20, 40, 60 second pre-buffering durations, respectively.

When MSPlayer enters the periodic re-buffering phase, we investigate how quickly it refills the playout buffer and compare its performance to that of other commercial players with HTTP range requests using default chunk sizes of 64 KB (Adobe Flash) and of 256 KB (HTML5) over single path WiFi and LTE \cite{rao2011network}. Similarly, we also look at different re-buffering sizes for 20/40/60 seconds.

\begin{figure}[t]
\raggedright
\includegraphics[scale=0.48]{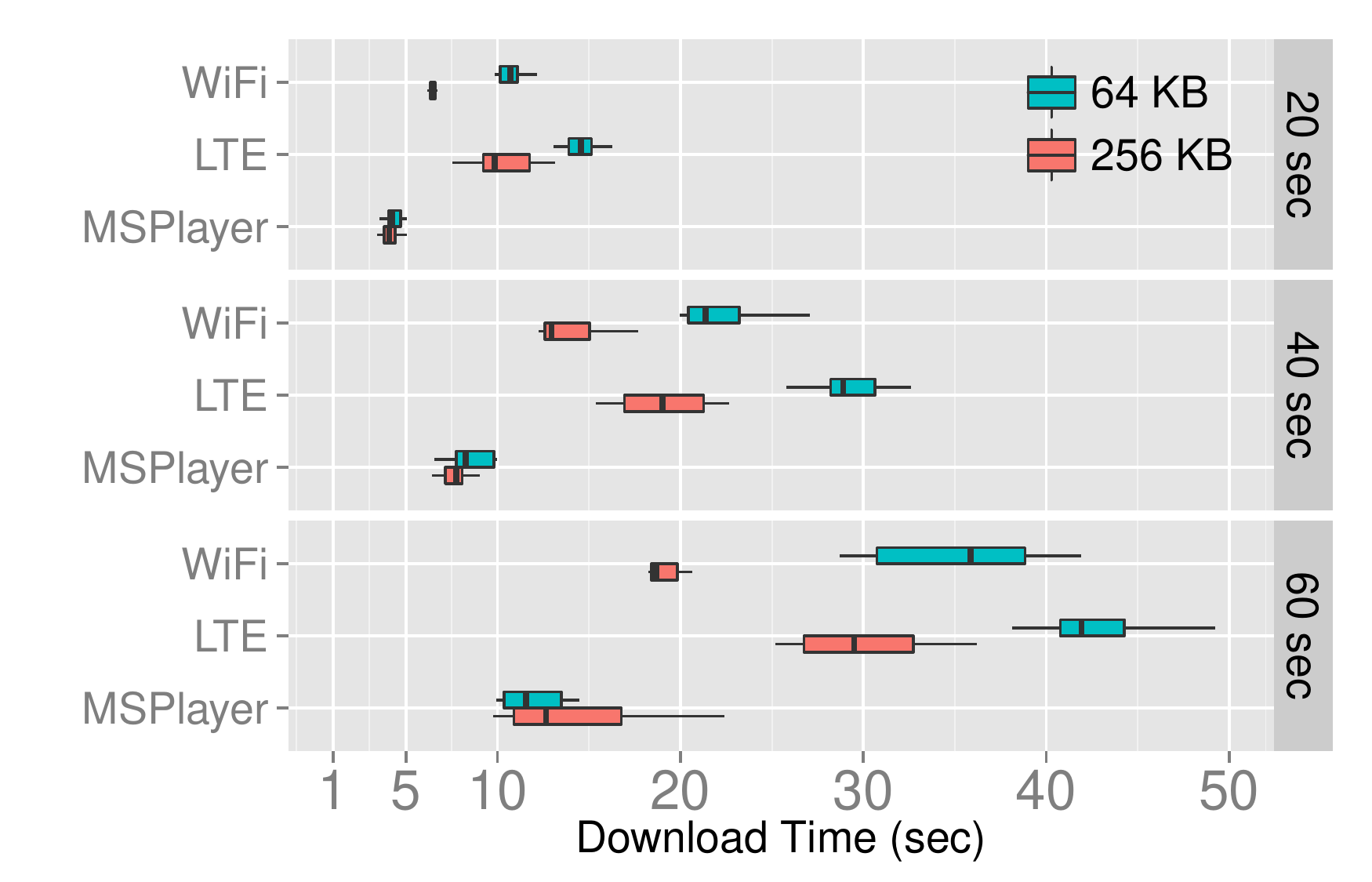}
\caption{Re-buffering 20/40/60 second video with HTTP byte range of sizes 64/256 KB for single-path WiFi, LTE, and MSPlayer over YouTube service.}
\label{fig:re_buf}
\end{figure}

Fig. \ref{fig:re_buf} presents the performance when streaming YouTube videos over single-path WiFi/LTE with HTTP byte ranges of sizes 64 KB and 256 KB for different re-buffering sizes. All of the players refill the playout buffer quickly when using larger chunks. This is because more requests are required for smaller chunks and introduces more overhead. MSPlayer, on the other hand, efficiently estimates network bandwidth and adjusts the chunk size accordingly. It outperforms all the single-path schemes and can significantly reduce the time to refill the playout buffer.

\begin{table}[h]\small
\newcolumntype{d}[1]{D{.}{\cdot}{#1} }
\renewcommand{\arraystretch}{1.2}
\centering
\caption{Fraction of Traffic over WiFi (mean \textpm std)}
\vspace*{1em}
\begin{tabular}{@{}ld{3.1}@{}ld{3.1}@{}ld{3.1}@{}l@{}} \toprule
   & \multicolumn{2}{c}{Pre-buffering } & \multicolumn{2}{c}{Re-buffering } \\ \cmidrule{2-5}
20 sec & 64.1&\textpm 9.3\% & 61.8 &\textpm 7.1\%  \\
40 sec & 60.1 &\textpm 15.0\% & 61.7 &\textpm 11.5\%   \\ 
60 sec & 63.7 &\textpm 12.6\% & 56.5 &\textpm 11.6\%  \\ 
\bottomrule
\end{tabular}
\label{tab:fraction}
\end{table}

In order to understand how the MSPlayer chunk scheduler distributes traffic over paths, we investigate the fraction of traffic carried by each path. Table \ref{tab:fraction} lists the fraction of traffic carried by WiFi for both pre-buffering and re-buffering phases with initial chunk size 256 KB. We observe that the WiFi path on average carries more than 60\% of traffic in the pre-buffering phase. This is mainly due to the fact that our design allows the fast path to start fetching video chunks as soon as it decodes necessary information from YouTube's web proxy server (as discussed in Sec. \ref{sec2:msmp}). In our experiments, the RTTs of the LTE network are two to three times larger than those of the WiFi network, and hence the WiFi path starts the streaming process earlier than the LTE path. 

During the re-buffering phase, the WiFi path slightly dominates packet delivery. This is because each path needs to wait for one RTT before receiving the first packet from the associated video source for each range request. As WiFi exhibits much smaller RTT values in our experiments, the WiFi path saves a time of length $R_2-R_1$ for each range request and introduces less overhead when compared with the LTE path in the re-buffering phase.

\section{Conclusion and Future Work}
\label{sec:future}

We propose a client-based video streaming solution, MSPlayer, that streams videos from multiple YouTube video servers via two interfaces (WiFi and LTE) simultaneously. MSPlayer manages to reduce video start-up delay and can quickly refill the video playout buffer for just-in-time high quality video delivery. It does not require kernel modifications at either the server or the client side. Moreover, it provides robust data transport and does not suffer from middleboxes in the networks as does MPTCP. Due to space constraint, we do not report the results on how MSPlayer provides robustness for video delivery in mobile scenarios.


As for future work, our scheduler currently does not take into account energy constraints when leveraging multiple interfaces on mobile devices \cite{huang2013depth}. Also, we use a simple periodic downloading mechanism for playout re-buffering. A more careful investigation of periodic downloading and ON/OFF mechanisms will be explored. Last, as we have taken an initial step to demonstrate the possibility of leveraging multiple video sources with different interfaces/paths in a real video service network, we only focus on using a constant video bit-rate. As dynamic adaptive streaming over HTTP (DASH) \cite{stockhammer2011dynamic} is now widely used, exploring how rate adaption can be integrated with MSPlayer and how MSPlayer can be used for other streaming services are also our future works.

\section*{Acknowledgments}
This research was sponsored by US Army Research laboratory and the UK Ministry of Defense under Agreement Number W911NF-06-3-0001. The views and conclusions contained in this document are those of the authors and should not be interpreted as representing the official policies, either expressed or implied, of the US Army Research Laboratory, the U.S. Government, the UK Ministry of Defense, or the UK Government. The US and UK Governments are authorized to reproduce and distribute reprints for Government purposes notwithstanding any copyright notation hereon. 

\newpage
\bibliographystyle{abbrv}
\bibliography{youtuber}

\end{document}